# Procedural Modeling of Urban Land Use

Tom Lechner[1],
Benjamin Watson[2]
Dept. EECS, Dept. CS
Northwestern U., NC State U.

Uri Wilensky[1],
Seth Tisue[1]
Program Learning Sciences
Northwestern Univ.

Martin Felsen[3],
Andy Moddrell[3]
School of Architecture
Illinois Inst. Technology

Pin Ren[1],
Craig Brozefsky[1]
Dept. CS, Prog. LS
Northwestern Univ.

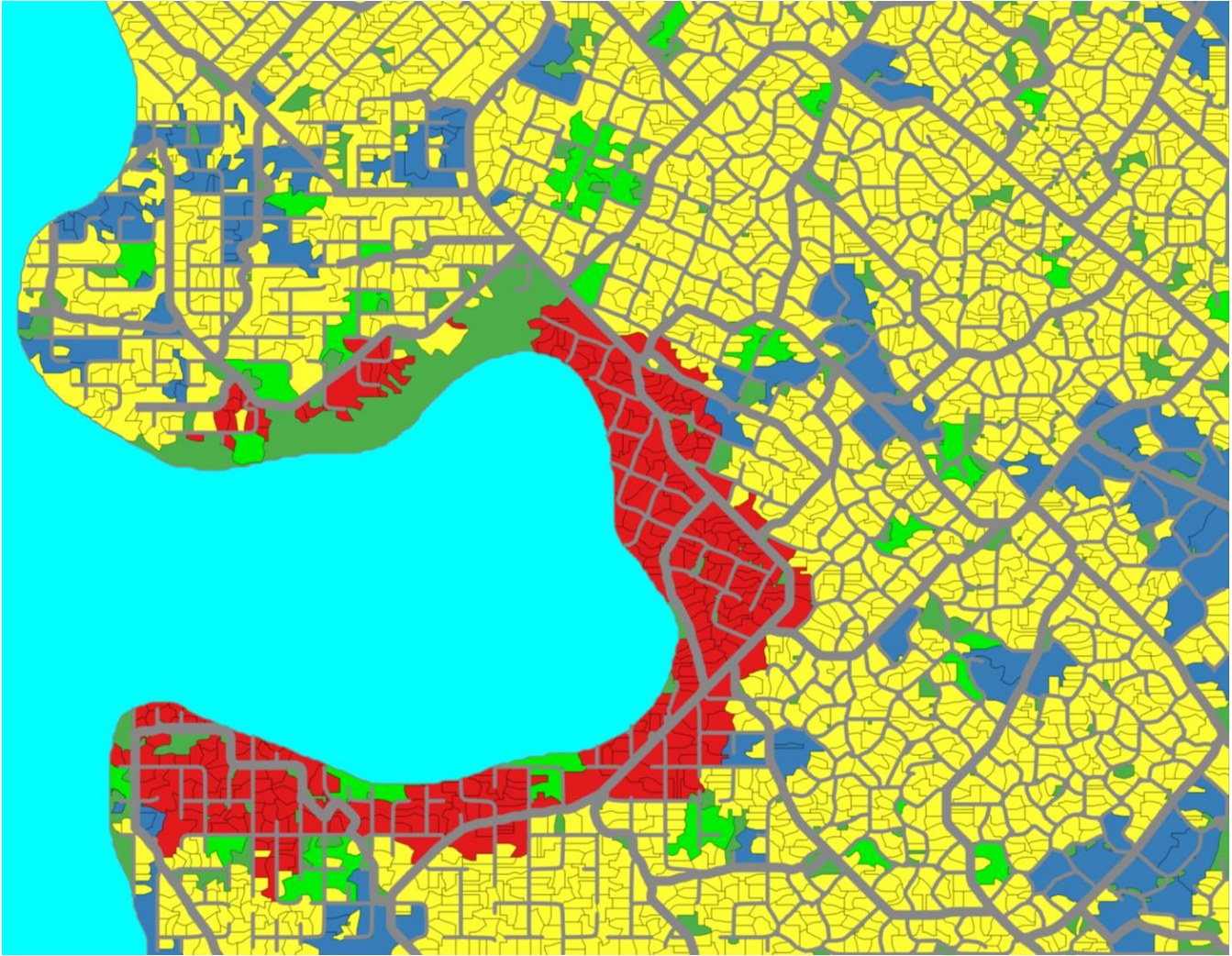

**Figure 1:** Vectorized output from our procedural city model at a three-mile width, showing residential (yellow), commercial (red), industrial (blue) and park (light green) development. The artist has painted "honey" near the bay to create an unusually dense commercial zone, and left parcel boundaries quite loose. Note the neighborhoods with differing road layouts.

## Abstract

Cities are widely used as content in digital productions, and their complexity makes them very difficult for artists to model. The few tools that help artists in this work do not model land use, meaning artists must arrange the buildings in the cities they create manually. We describe a method for procedurally generating typical patterns of urban land use using agent-based simulation. Given a terrain description, our system returns a map of residential, commercial, industrial, recreational park and road land uses, including age and density of development. Artists can interact with the map via a painting interface to establish global developmental behavior, guide local development trends, or directly set desired land use. Our results conform to modern patterns of land use, but each generated city is unique.

CR Categories: I.6.5 [Computing Methodologies]: Simulation and Modeling — Model Development; I.3.5 [Computing Methodologies]: Computer Graphics — Computational Geometry and Object Modeling; J.5 [Computer Applications]: Arts and Humanities — Architecture.

Keywords: cities, urban planning, urban design, urban development, urban geography, procedural modeling, agent-based simulation, behaviors, complexity.

## 1 Simulating land use

Cities are important elements of content in digital productions, but their complexity and size make them very challenging to model. Few tools exist that can help artists with this work, even as rapid improvements in graphics hardware create demand for richer content without matching increases in production cost.

We propose a method for procedurally generating realistic patterns of land use in cities, automating placement of buildings and roads for artists. Given a geographical map, our current system models the development of five fundamental types of land use: residential, commercial, industrial, roads and parks. Artists can allow development to proceed completely autonomously, adjust global developmental behavior, or interact with the evolving map of land use to guide or manually set local development. The system outputs a map of land use including property lines, age and density of development, and population. Each generated city is unique, but corresponds with modern patterns of land use. Figure 1 shows one artist-steered example.

## 2 Procedural modeling and motivation

Computer graphics has long been used to assist architects and urban planners, primarily through interactive modeling and visualization. Yet research in support of urban planning has only recently begun, with effort concentrating on capture [Teller et al. 2003; Frueh and Zakhor 2003] and interactive walkthrough of entire cities [D´ecoret et al. 2003; Schaufler et al.2000].

Computer graphics researchers have also dedicated considerable effort to assisting artists in their work. One particularly relevant stream of this research develops highly automated (procedural) modeling tools [Ebert et al. 2002]. To date most of this work has modeled complex natural objects and phenomena, including among many others fire and explosions [Reeves 1983], plants [Deussen et al. 1998], and flocking and schooling [Reynolds 1987]. Effort on procedural modeling of human artifacts has been relatively sparse, including work on modeling tilings [Miyata 1990; Legakis et al. 2003] and truss structures [Smith et al. 2002].

Recently researchers in procedural modeling have begun addressing applications in architecture and urban planning, motivated by many of the same problems we focus on in our work. Even with the best capture techniques, urban models often contain gaps or lack detail. Lewis and Séquin [1998] ameliorate this problem by generating 3D building models from 2D plans. Given approximate 3D building geometries, Wonka et al. [2003] generate detailed facades using shape grammars, with results exhibiting cultural variety and responding to the influences of population and material. Other researchers have focused on procedural modeling of road networks. Parish and Müller [2001] grow these networks using L-systems, given input terrain and population maps. Users of their systems can specify that roads in certain locations use certain layouts (e.g. gridded and radial).

The computer game SimCity [Electronic Arts 2003] allows players to act as city planners, creating road and land use maps manually and then watching simulated cities evolve according to these maps. Our project has benefited from a close collaboration with the SimCity developers at Electronic Arts.

Our contributions to this research endeavor are many:

*Land use.* Our system simulates the distribution of land uses in urban landscapes, automating the placement of buildings and structures in digital cities for artists and greatly increasing realism.

*Steerability.* Because our tool uses agent-based simulation rather than L-systems, artists can pause and alter the course of the simulation freely to meet applied needs, without concern for whether or not the change is a valid construct of an output language.

*History.* Our system simulates the continuous history of urban development. This enables the system to assign different ages to various parts of the urban landscape, greatly improving variety and realism.

## 3 Simulating urban development

Long before the procedural city modeling efforts described above, urban planners, designers and geographers built their own simulations. Planning simulations [Brail and Klosterman 2001] are focused on predicting near-term futures for existing cities, and therefore rely heavily on input from geographic information systems (GIS) and census data. In addition, because planners are much more interested in global statistics than fine spatial detail, most planning simulations only output statistics at global or extremely large spatial scales. One notable exception is Kwartler et al.'s CommunityViz [Kwartler and Bernard 2001], which uses agent-based based modeling, the same technique used by Reeves [1983] for his particle systems and by Reynolds for his boids [1987]. CommunityViz simulates development in small towns, outputting detailed spatial maps while still relying on extremely detailed GIS input.

Simulators built by urban designers are used to envision possible urban improvements and futures [Bettum and Hensel 2000; Testa et al. 2000]. These simulations also rely heavily on GIS input, and since they are primarily tools for conceptual design, their output is extremely abstract and symbolic.

Because they want to understand the basic processes of urban development at multiple scales, the needs of urban geographers are closest to our own, and we have learned much from their work. Geographic simulations are carefully validated and quite realistic. However until recently, urban geographic simulations were also created only at extremely large scales [Benenson and Torrens 2004]. Even newer simulations [Clarke et al.1997; White and Engelen 2000] are often too large in scale to enable detailed positioning of individual buildings and streets. Regardless, our goals (supporting artists) and those of geographers (understanding the real world) are fundamentally different, and result in fundamentally different solutions. Geographic simulations require heavy input, focus on modeling of realistic processes and once started, are usually unsteerable. In contrast, artists need solutions that require little input, focus on realistic *output* and are as interactive as possible.

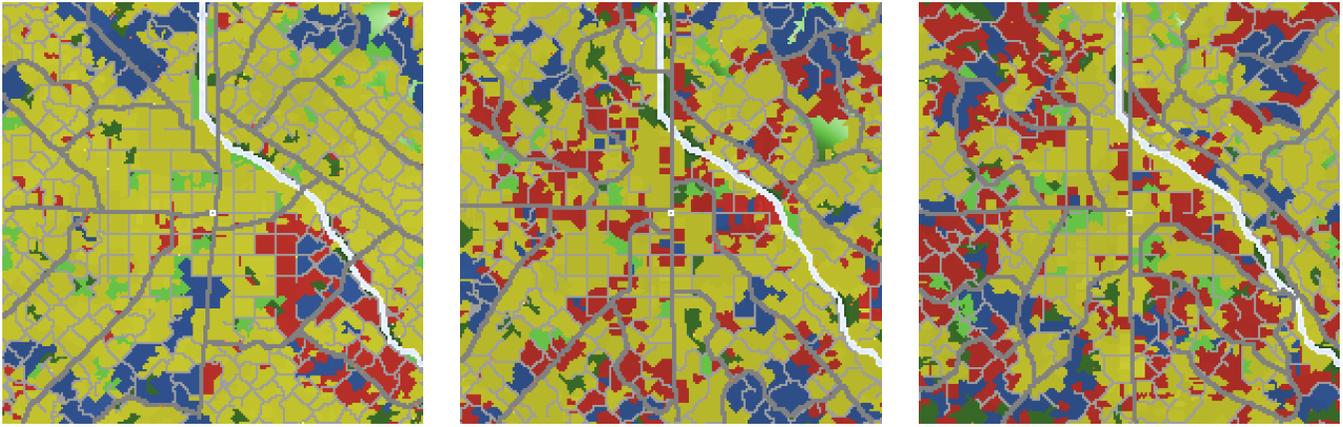

**Figure 2:** The artist increases red, commercial land use from left to right by manipulating the land cover constraints $\lambda_t$.

## 4 Patterns of urban development

According to urban experts, the most salient components of the urban development are residential, commercial, industrial and road development, which together result in roughly 80% of urban land use (40%, 5%, 10% and 25% for each respective land use) [Eisner et al. 1993]. We therefore focus primarily on these land uses in our work.

Each property developer seeks the regions most appropriate for its use [Eisner et al. 1993; Kostof 1991; Alexander et al. 1977]. Residential developers prefer to place the homes they build in large clusters, and away from busy roads and industry. They favor waterfronts, above-average elevation for views, and parks. Residential developments also use the smallest and least densely populated parcels.

Commercial developers place their stores and offices close to their markets and transportation corridors. For this reason these developments cluster as well, but this tendency is counteracted by the limited size of local markets. These developers also favor relatively flat terrain. Commercial development parcels are larger than residential parcels, and much more densely populated.

Industrial developers prefer flat land and access to transport. The heavy traffic and pollution they generate make them uncomfortable residential and park neighbors. Industrial parcels are the largest, but less densely populated than commercial parcels.

Unfortunately, parks are typically an afterthought in cities, with large parks not appearing until the city is quite mature. For this reason, parks most often appear in neglected regions of the cityscape, such as flood plains and rugged terrain. As amenities for the home rather than the work life, parks are generally closer to residential than industrial areas.

Roads are organized into a hierarchy, from freeways to primary roads down to tertiary roads [Eisner et al. 1993; Alexander 1977]. Higher levels of the hierarchy are wider roads serving high-speed travel over long distances. Lower levels provide access from homes and places of work to higher levels of the hierarchy, and therefore are smaller and only support travel at low speeds. Road networks are often organized in tight or loose grid patterns, or alternatively in more "organic" patterns that respond to historic paths, property lines and topography [Kostof 1991; Carmona et al. 2003]. Our system currently models both primary and tertiary roads, and allows artists to specify where these roads should be organized organically or into grids.

## 5 Procedural modeling of land use

We implement our simulation in NetLogo, a programming language and agent-based simulation environment developed by Uri Wilensky [1999]. The high-level nature of the language enables rapid development, aids low-cost experimentation with different solutions, and facilitates our collaborations with architects and urban planners with its transparent, "open box" design.

We associate each land use with distinct *developer* agents, which act in a simulated environment we call the *world*. The world consists of a rectangular grid of atomic areas called *patches*, which store local state. *p* refers to an agent's current patch, and makes the components of the patch's state accessible as fields (e.g. p.e is the patch's elevation). Each patch is 40 ft (~12m) square, allowing the simulation to represent tertiary roads using a one-patch width. With a 200x200 patch world, our system then models an environment approximately 1.5 miles square.

### 5.1 Input, interaction and output

The primary source of input to our simulation is a terrain height map that can either be input directly using a painting interface, or read in from a file. In addition, artists may shape urban development using a number of parameters (though serviceable cities are generated using parameter defaults). Figure 1 shows the effect of road layout parameters, with gridded regions in different orientations, and "organic" ungridded regions. Figure 2 shows how artists can adjust the proportions of residential, commercial, industrial and park development. In Figure 3, the artist varies the predominance of the urban center, with density of development clustering tightly or loosely. Finally in Figure 4, the artist creates cities in which different land uses cluster more or less tightly. Some parameters (e.g. proportions of development types) are global and affect the entire city, while others (e.g. road layout) are local, with artists applying them to a specific neighborhood using the painting interface.

This initial artist input is useful, but the demands of digital production are often much more specific. Artists may therefore steer the simulation by removing unsatisfactory development, adding desired development, and adjusting simulation parameters. These actions may be taken at any point during the development of the city. One particularly useful parameter is *honey*, which enables artists to add local incentives encouraging certain types of development. In Figure 6, the artist demolishes a residential neighborhood and paints it with honey encouraging industrial development. Another *reserved* parameter prohibits any development. Students in one of our architecture courses used this parameter to recreate the neighborhoods of Madrid and Berlin shown in Figure 7.

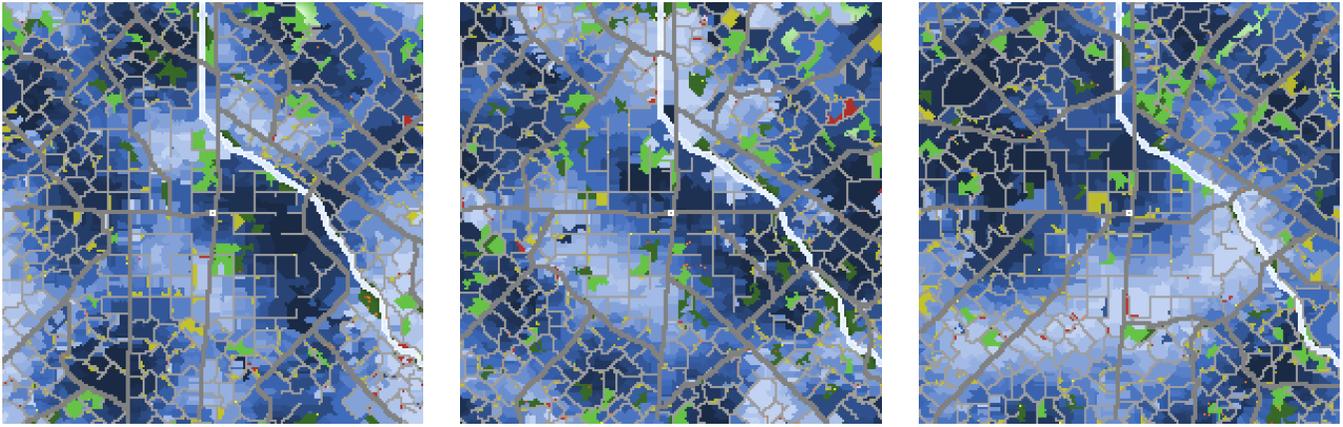

**Figure 3:** The artist increases population clustering, from left to right, by adjusting the density smoothness constraints $\sigma_t$. This effectively strengthens the urban core. In these density views, higher population density is indicated by whiter parcels, and use is not displayed.

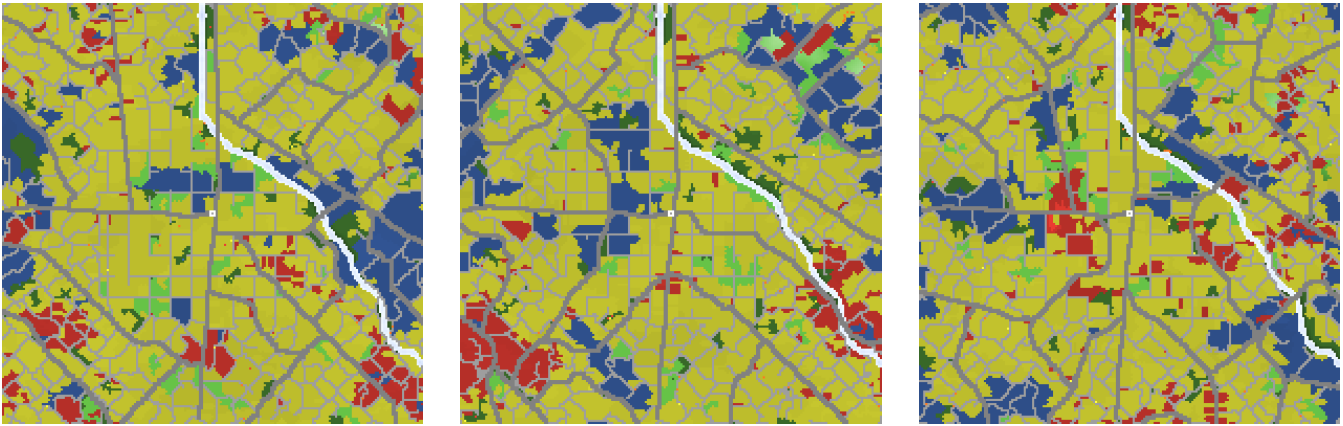

**Figure 4:** The artist varies the clustering behavior of residential, commercial and industrial developers by increasing attribute weights from left to right. Because red commercial developments have fewer proximity constraints than residential and industrial uses, they respond most strongly, forming fewer, denser (brighter red) clusters.

Our system generates a rasterized map of land uses, with residential, commercial, industrial and park uses occupying properties called *parcels* made up of several patches, and roads occupying series of patches connected to form a transportation network. Each parcel describes its population, age, and value, with parcel density the ratio of parcel population over parcel patches. Roads are identified as either primary or tertiary. These layouts may then be populated with appropriate 3D buildings, as illustrated in Figure 17.

### 5.2 Property developers

*Property developer* agents construct parks as well as residential, commercial and industrial buildings. Each of these uses is matched to a corresponding agent type. As Figure 5 shows, during each simulation tick a property developer prospects, builds and checks profitability.

```
for each simulation tick
  devSites = prospect(type,devSites)
  for each site in devSites
    newDev = build(site,type)
    if profitable(site,type,newDev) then commit(newDev)
  endfor
endfor
```

**Figure 5:** High-level behavior of property developers.

Figure 8 outlines prospecting behavior, which consists of moving to a new location, and then updating the agent's list of developable sites devSites. The new location is the most valuable member of devSites in the developer's current local area circle(5) (the diameter 11 circle centered at the developer). If circle(5) contains no developable sites, or if the agent has not created a profitable development recently, it relocates globally to one of the most valuable fifth of all developable sites, and resets its list of recently seen undeveloped patches devPatches.

devSites contains parcels or empty patches that are adjacent to a road and not reserved by the artist. Parcels must also intersect circle(5) and not be prohibited for development by the developer's type: residential use cannot be converted directly to industrial use (and vice versa), while parks may not be converted to any other use, and may only be installed in undeveloped land. Empty patches may not be in circle(5). Instead, the developer maintains a list devPatches that is a subset of devSites, and after each move, adds all empty patches in circle(5) to devPatches, and then removes the least valuable 10% from the resulting set.. This non-local behavior increases selectivity in locating new development.

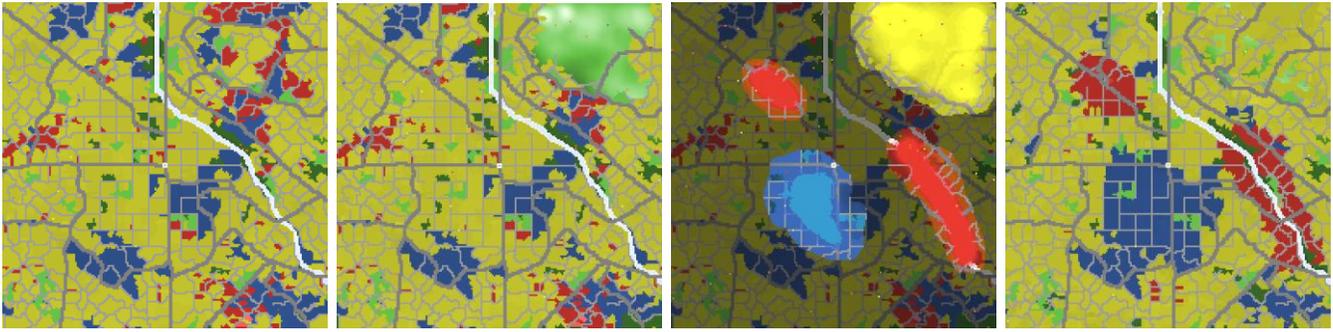

**Figure 6:** The artist steers the urban simulation. Left, the state of the simulation before editing. Left of center, the artist erases an undesirable commercial and industrial cluster. Right of center, the artist paints residential, commercial and industrial honey on the city to attract desired uses to certain neighborhoods. Right, the resulting city, with roads and parcels automatically placed.

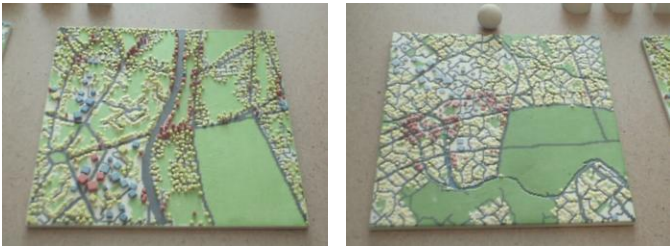

**Figure 7:** Architecture students used the simulation to recreate neighborhoods in Madrid (left) and Berlin (right). They produced 3D models of their results with a rapid prototyping machine.

```
prospect(type,devSites)
 if devSites ≠ ∅ and (recentCommit or recentRelocate) then
  move to select(devSites ∩ circle(5),max(value))
 else begin // relocate globally
  wDevSites = select(world,parcels_convertible_to(type))
  wDevSites = wDevSites ∪ select(world,empty)
  move to random(select(wDevSites,top5th(value)))
  devPatches = ∅
 endif
 devParcels = select(circle(5),parcels_convertible_to(type))
 devPatches = select(devPatches ∪ select(circle(5),empty),top90(value))
 devSites = devParcels ∪ devPatches
 return validSites
```

**Figure 8:** Prospecting behavior of property developers.

A property developer can build by improving an existing site already put to its own use, by converting a site to its own use, or by forming a parcel on undeveloped land. Figure 9 shows that to improve a site, a developer simply increases the site's population by 1. To convert a site, the agent changes the site's type.

```
build(site,type)
 if site is patch then newDev = newParcel(buildLot(site,type),type)
 else if site is parcel then begin
  newDev = copy(site)
  if site.type ≠ type then newDev.type = type
  else newDev.population = site.population + 1
 endif
 return newDev
```

**Figure 9:** Building behavior of property developers.

To form a residential, commercial or industrial parcel out of undeveloped patches, a developer chooses a random target parcel size from its type's size range. Starting from the only empty patch currently in the site, the developer moves perpendicularly away from the road, adding each newly reached patch to the parcel. When the developer has traveled half the block length $B$, it stops moving away from the road and begins widening the strip of patches now in the parcel by adding patches along either side of the strip, until the target number of patches is in the parcel, or no undeveloped patches are adjacent to the parcel. If the resulting parcel is smaller than the minimum in the agent's size range, it attempts to merge the parcel with adjacent parcels with the same use, without surpassing the range's maximum. If the resulting parcel is in the type's size range, the developer records the creation date in the parcel and assigns it a population proportional to its type's minimum density. Otherwise the developer aborts parcel creation.

Parks are built by the public sector, and are subject to additional artist-defined constraints. For park building to proceed, cities must have a certain minimum population and contain less than a certain number of park patches both per resident and per developed patch. To form a park, the developer chooses a random target size from its size range. It then scales this size by the value of the site and the size of the city, to create bigger parks at particularly valuable sites, and in bigger cities. The developer then starts a flood fill from the site's only empty patch. From the set of empty, unreserved patches adjacent to the current parcel, the agent adds patches more valuable than half of the other patches in the set until the parcel reaches the target size, or is completely surrounded by development. If the flood fill stops before reaching the minimum park size, the park developer aborts parcel creation.

To improve honey's effectiveness, a developer may immediately swap an existing parcel for a honeyed parcel. Without this mechanism, simulation constraints on the global proportions of uses hinder the ability of developers to take advantage of honey, much like real world economies limit the effectiveness of government incentives. The existing parcel in the swap must be adjacent to a parcel of a different type, and will be the least valuable of such parcels. The swap is executed only if the honeyed parcel is more valuable than this existing parcel. At completion of the swap, the existing parcel reverts to an undeveloped state.

To check profitability, the developer produces a cost-profit model by in fact developing on each site in devSites and calculating the change in land value. As Figure 10 illustrates, if the site's value would increase by at least $P$%, it would be profitable. If the site was undeveloped or its use would not change, then this profitability commits the agent to the new development. If the site's use would change, then the profit from the new development must be greater than any loss in value to developers for the prior use. This helps ensure that sites are developed for the most profitable use, and brings some stability to the urban market.

**Table I:** Attributes used to value land. p is the current patch, with p.d$_p$ distance to primary roads, p.d$_w$ distance to water, p.e elevation and p.population development capacity. The size field is the number of patches in a parcel. $\bar{e}$ is average elevation, $e_{offset}$ is desirable height above $\bar{e}$, and $e_w$ is water elevation. Untransformed attributes not in the table include honey as well as p.d$_{pk}$ and p.d$_{com}$, the distances from p to the nearest park and commercial development.

| Attrib | Description | Equation |
|---|---|---|
| $e_h$ | Elevation advantage | $e^{\frac{(p.e-\bar{e}-e_{offset})^2}{-128}}$ |
| $e_v$ | Variance in elevation (negative) | $e^{-setVar(circle(5).e)}$ |
| $e_{pv}$ | Variance in elevation (positive) | setVar(circle(5).e) |
| $e_{fp}$ | Flood plain elevation | $(p.e-e_w)^{-2}$ |
| $d_{pr}$ | Proximity to primary roads | $e^{-p.d_p}$ |
| $d_w$ | Proximity to water | $(1+p.d_w)^{-2}$ |
| $d_m$ | Proximity to market | $0.5(1+d_{pr})c_c d_c$ |
| $d_r$ | Residential density | $\frac{setSum(res\_parcels\_in(circle(5)).population)}{\|circle(5) \cup patches\_in(parcels\_in(circle(5)))\|}$ |
| $d_c$ | Commercial density | $\frac{setSum(com\_parcels\_in(circle(5)).population)}{\|circle(5) \cup patches\_in(parcels\_in(circle(5)))\|}$ |
| $d_i$ | Industrial density | $\frac{setSum(ind\_parcels\_in(circle(5)).population)}{\|circle(5) \cup patches\_in(parcels\_in(circle(5)))\|}$ |
| $c_c$ | Commercial clustering | $1/max(1 \cup neighbors(parcels\_in(p)).size)$ |
| $x$ | Anti-worth | $(1/v_r + 1/v_c + 1/v_i)$ |

**Table II:** Weight vectors applied to land attributes by property developers. All attributes are normalized with μ.

| Weights | $e_h$ | $e_v$ | $e_{pv}$ | $e_{fp}$ | $d_w$ | $d_r$ | $d_i$ | p.d$_{pk}$ | $d_{pr}$ | $d_m$ | p.d$_{com}$ | $x$ |
|---|---|---|---|---|---|---|---|---|---|---|---|---|
| $W_r$ | .3 | | | | .3 | .4 | | | | | | |
| $W_c$ | | .2 | | | .15 | .15 | | | | .4 | .1 | |
| $W_i$ | | .5 | | | .3 | | .1 | | .1 | | | |
| $W_p$ | | | .1 | .1 | .1 | .1 | | .4 | | | | .2 |

```
profitable(site,newDev)
 if newDev = ∅ then profit = false
 else begin
  profit = value(newDev.type,newDev)/value(newDev.type,site) ≥ (1+p/100)
  if (site is parcel) and (site.type ≠ newDev.type) then begin
   prevLoss = min(0,value(site.type,newDev) - value(site.type,site))
   newGain = value(newDev.type,newDev) – value(newDev.type,site)
   profit = profit and prevLoss ≤ newGain
  endif
 endif
return profit
```

**Figure 10:** Profit-checking behavior of property developers.

*5.2.1 Determining Value*

The values that developers assign to properties are the most important factor in determining their behavior, and the primary means through which artists can influence it. Property developers of type $t$ calculate patch value $v_t$ using the following equation:

$$v_t = \pi_t \lambda_t \delta_t \sigma_t (W_t \cdot A) + h_t$$

where $A$ is a vector of attributes describing the world in the locality of the patch, $W_t$ is the vector of importance weights given to each of those attributes by developers of type $t$, $\pi_t$ is the proportional population constraint, $\lambda_t$ is the proportional land cover constraint, $\delta_t$ is the proximity constraint, $\sigma_t$ is the density smoothness constraint, and $h_t$ is honey in the range [0,1]. Note that when patches have been developed and assigned to a parcel, value is determined not per-patch but per-parcel, with that value being the mean of the constituent patch values.

Table I details the attributes in $A$, which measure a qualities that developers consider valuable or attractive. Since these attributes have different ranges and units of measurement, they are normalized by the function $\mu$:

$$\mu(x, \bar{x}) = \begin{cases} 2^{(1-\bar{x}/x)} : x \leq \bar{x} \\ 2 - 2^{(1-x/\bar{x})} : x > \bar{x} \end{cases}$$

where $x$ is the value of an attribute of a particular site (always positive) and $\bar{x}$ is the mean of all such values in the city. $\mu$ limits attribute values to the range [0,2]. If the value and its mean are equal, $\mu$ is one. As the ratio of an attribute value to its mean increases, $\mu$ approaches two, while as the ratio drops, $\mu$ approaches zero.

$W_t$ indicates the importance property developers attach to each attribute when valuing a patch. Table II shows the default weights. These weights sum to 1, so that a patch that is average in every respect will have an unconstrained value of 1. Figure 4 demonstrates the effects on clustering by use produced by manipulating some of these weights. The constraints listed in Table III implement powerful global restrictions on property development. The population constraint $\pi_t$ maximizes value when the population occupying land dedicated to use $t$ is a certain proportion of the entire city's population. Similarly, the land coverage constraint $\lambda_t$ maximizes value when the number of patches dedicated to use $t$ is a certain proportion of the number of patches in the city. Figure 2 shows how the portions of the city dedicated to each type change as these constraints are varied. The constraint $\sigma_t$ controls the smoothness of changes in population over space for each land use $t$. In Figure 3, changes in this constraint vary the strength of the city's core. Finally, the proximity constraint $\delta_t$ discourages certain types of developers from developing too close to others: in particular, residential and industrial uses are rarely located close to one another, while parks are rarely located close to industrial and commercial developments.

### 5.3 Road developers

Three types of developers build roads: *tertiary extenders*, which ensure that undeveloped land is accessible; *tertiary connectors*, which ensure that the tertiary road network is adequately interconnected; and *primary developers*, which ensure that residents can move through the city as a whole quickly.

All road development is shaped by several local, artist-paintable constraints. $g_x$ and $g_y$ define the local grid spacing in two dimensions. $g_\theta$ allows the grid to be rotated locally in the world's plane. $g_{dx}$ and $g_{dy}$ indicate local tightness in each dimension of the grid. If $g_{dx} = g_{dy} = 0$, road patches must be directly on the grid. If $g_{dx} \geq g_x$ and $g_{dy} \geq g_y$, the grid does not constrain road layout at all, so the local road network will be completely ungridded and "organic". Finally, $D_t$ limits road density d$_{road}$, the number of road patches within each local neighborhood circle(5). Figure 1 shows the impact of varying these parameters within a single city.

**Table III:** Constraints used to value land. $p_t$ and $l_t$ are total population and land cover for use (type) $t$, $p_C$ and $l_C$ are city-wide population and land cover, while $p_{Dt}$ and $l_{Dt}$ are desired proportions of $p_t$ to $p_C$ and $l_t$ to $l_C$. $d_{ti}$ is the initial density of a building during construction. For parks, $\pi_t = \lambda_t = \sigma_t = 1$. All attributes are again normalized with μ.

| Constr | Description | Equation |
|---|---|---|
| $\pi_t$ | Population constraint for use $t$ | $max\left(0, 1 - \left(0.9\left(\frac{p_t}{p_C p_{Dt}} - 1\right)\right)^2\right)$ |
| $\lambda_t$ | Land cover constraint for use $t$ | $max\left(0, 1 - \left(0.9\left(\frac{l_t}{l_C l_{Dt}} - 1\right)\right)^2\right)$ |
| $\delta_r$ | Proximity constraint for residential uses | $0.8 + 0.2(10^{-d_i})$ |
| $\delta_c$ | Proximity constraint for commercial uses | $0.8 + 0.2(10^{-p \cdot d_{pk}})$ |
| $\delta_i$ | Proximity constraint for industrial uses | $0.4 + 0.2(10^{-d_r}) + 0.4(10^{-p \cdot d_{pk}})$ |
| $\delta_p$ | Proximity constraint for parks | $0.7(10^{-d_i}) + 0.3(10^{-d_c})$ |
| $\sigma_t$ | Density smoothness constraint for use $t$ | $\text{clamp}(0, log_3(4.2 - d_{ti} / min(d_{ti} \cup \text{neighbors}(p).d_{ti})),1)$ |

Figure 5 shows the common high-level behavior shared by all road developers. During each simulation tick, a road developer prospects by moving to a new location. If that location meets current road constraints and needs a road, then the developer attempts to build a new road. If the resulting road meets certain criteria, the developer commits to this road. Otherwise the road is removed.

```
for each simulation tick
  move to newLocation(type)
  if (meetsConstraints(p) and needsRoad(p,type)) then
    roadSegment = buildRoad(p,type)
    if valid(roadSegment,type) then commit(roadSegment)
  endif
endfor
```

**Figure 11:** High-level road developer behavior.

### 5.3.1 Tertiary road extenders

Tertiary road extenders push access roads into empty territory, making it developable. An extender prospects by hill-climbing through a landscape defined by distance to the road network roadDist, which is stored at each patch. It will not move more than $D_{max}$ patches from the road network. To avoid local maxima, an extender does not revisit recently examined patches. If despite this a local maximum is reached, or if an extender has prospected for too long without attempting to build, it transports itself to a distant patch that meets local road constraints and needs a tertiary extension.

```
buildRoad(start,extension)
  possRoad = empty_list; nextPatch = {start}
  while (not road(p) and ||nextPatch|| > 0)
    move to select(nextPatch,random)
    possRoad = append(roadSeg,p)
    nextPatch = select(neighbors(p),meetsConstraints())
    nextPatch = select(nextPatch,min(roadDist))
    nextPatch = applyTiebreakers()
  endwhile
  return possRoad
```

**Figure 12:** Building behavior of tertiary road extenders.

Any location at least $d_{min}$ patches from the road network needs a road, and an extender will attempt to build a road to it, if the location meets local constraints. Extender building behavior is illustrated in Figure 12. To find the road's path, an extender descends the same roadDist landscape it climbed when prospecting, checking that each patch it reaches fulfills the road constraints. Given a choice between two or more patches, an extender uses two tiebreakers. First, it will choose the patch that is on a parcel boundary. If that does not resolve the choice, the extender chooses the patch with the lowest absolute change in elevation. Otherwise, the choice is random. When the proposed road extension is complete, an extender will reexamine it to ensure that the road density constraint $D_t$ is still met. If so, it commits to the road and adds it to the city.

### 5.3.2 Tertiary Road Connectors

Connector agents ensure that the tertiary road network is adequately interconnected, enabling fairly direct travel between any two points. A prospecting connector moves along the road network, choosing a random direction whenever it reaches an intersection. At each prospected patch p, the connector chooses a random destination road patch dest within the radius $r_c$ of p. p needs a road if the shortest path on existing roads to dest is too long, and the region along the direct line between p and dest contains no roads. A path is too long if it is $c_{ratio}$ times longer than dist(p,dest), the actual distance between p and dest, or if any point on that path is beyond the radius $r_c$.

When building a road, a connector begins linking start and dest by moving toward dest as outlined in Figure 13. Each new road patch nextPatch must fulfill road constraints. When more than one patch q meets the constraints, the connector chooses a nextPatch that heads toward dest while avoiding existing roads and changes in elevation. It does this by minimizing the cost function $c_{tc}$:

$$c_{tc} = 0.2n(q.d_{road}) + 0.1(1 - n(q.\text{roadDist})) + n(\text{dist}(q,\text{dest})) + 0.04n(|q.e - \text{dest}.e|)$$

where $n(z) = z/(max(z), \forall$ q). Because connectors work in a highly constrained, already developed environment, they can backtrack a few times should an initial road path prove unsuccessful.

```
buildRoad(start,connection)
  prevStates = empty_stack[3]; possRoad = empty_list; nextPatch = {start}
  repeat
    repeat
      move to nextPatch; possRoad = append(possRoad,p)
      possPatches = select(select(neighbors(p),meetsConstraints()),min(c_rc))
      nextPatch = select(possPatches,random)
      push(prevStates,possRoad,possPatches-nextPatch)
    until (road(nextPatch) or ||nextPatch|| = 0)
  until not restoreState(pop(prevStates))
  return possRoad
```

**Figure 13:** Building behavior of tertiary road connectors.

Building ceases when a connector reaches a road patch, whether or not it is dest. Because the destination may change in this way, before an agent commits to a proposed complete connection, it must confirm not only that the connection still meets the density constraint $D_t$, but also that it is $1/c_{ratio}$ times shorter than the shortest path between start and the connections's actual ending patch on the existing road network.

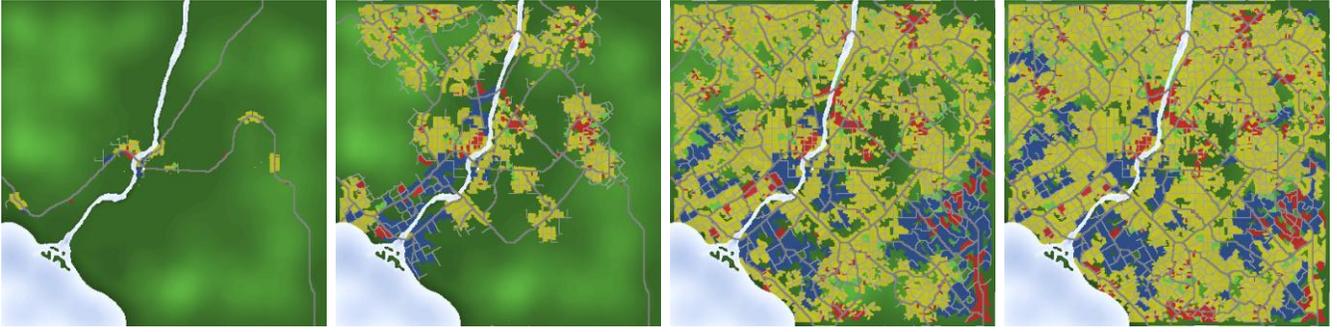

**Figure 15:** The developmental history of city nine miles square. Note the clustering at primary road intersections, and the early development along the shoreline. An industrial (blue) region develops away from the city center, but the city soon swallows it.

*5.3.3 Primary road developers*

Primary road developers connect the city's center to its surround and ensure that a primary road will usually be nearby. Primary roads need not adhere to gridding and road density constraints. Nevertheless, because they very often overlay the existing tertiary network, primary roads are still very much shaped by the constraints. To prospect, a primary developer hill-climbs through a $d_p$ (distance to primary road) landscape. To avoid overly dense primary road development, it will not enter the circle(5) neighborhood surrounding other primary developers. To avoid local maxima, the developer will not move to recently visited patches. Should it prospect without successfully building for too long, the agent transports itself to a developed patch far from the primary road network.

**buildRoad**(start,primary)
  nearDest = *select*(*world*,*primary* **and** min(roadPath(start)))
  *if* (rand(0,1) > .5) *then* farDest = *select*(*world*,*road* **and** *min*(dist(city-ctr)))
  *else* farDest = *select*(*world*,*edge* **and** oppositeDir(nearDest))
  primRoad = buildPrimary(start,nearDest)
  append(primRoad,buildPrimary(start,farDest))
return primRoad
**buildPrimary**(start,dest)
  newRoad = empty_list; nextPatch = {start}
  *repeat*
    newRoad = *append*(newRoad,*path*(*p*,nextPatch)); *move to* nextPatch
    *if* devInSlice() *then* nextPatch = *select*(neighbors(*p*),min(*c_{pu}*))
    *else* nextPatch = *select*(distantNeighbors(*p*),min(*c_{pr}*))
    nextPatch = *select*(nextPatch,random)
  *until* (primaryRoad(nextPatch) *or* edge(nextPatch) *or* ||nextPatch|| = 0)
**return** newRoad

**Figure 14:** Building behavior of primary road developers.

Tertiary road locations $d_{minP}$ patches along the network from the nearest primary road need a link to the primary network. A road developer builds such links using the behavior in Figure 14. It constructs two primary road segments, toward a nearby and a distant destination. The near destination is the primary road patch at end of the shortest path on the tertiary road network. The distant destination is either the road patch nearest the center of population density, or a patch at the edge of the world, in the opposite direction from the near destination.

A primary road developer builds roads in two modes: *urban* and *rural*. The developer is in urban mode when development exists in its *view slice*, the eighth of circle(5) neighborhood within 22.5 degrees of its current heading. In this mode, the constructed primary road avoids water, while heading toward dest as well as tertiary and especially primary roads. Candidate road patches q are neighbors of the last allocated road patch, and the developer chooses among them by minimizing the cost function

$$c_{pu}=[n(\alpha_{pu})\ n(\beta_{pu})\ n(\gamma_{pu})\ n(\delta_{pu})\ n(\varepsilon_{pu})\ n(\zeta_{pu})]\bullet[0.3\ 0.2\ 1\ 0.1\ 1\ 3],$$

where $\alpha_{pu}$ = dist(q,dest); $\beta_{pu}$ = distance of q from the line between start and dest; $\gamma_{pu}$ = 0 if road(q), 1 otherwise; $\delta_{pu}$ = 0 if q is a parcel boundary, 1 otherwise; $\varepsilon_{pu}$ = q.$d_p$; $\zeta_{pu}$ = $d_w^2$ times the developer's current heading dotted with the direction to water squared.

In rural mode the constructed road again avoids water and now also changes in elevation. It heads toward dest as well as primary and tertiary roads. Candidate road patches are organized into several paths, each defined by a ray emanating from p to a patch on the edge of the developer's view slice. The winning set of patches is chosen by minimizing the cost function

$$c_{pr}=[n(\alpha_{pu})\ n(\beta_{pu})\ n(\gamma_{pu})\ n(\delta_{pu})\ n(\varepsilon_{pu})\ n(\zeta_{pu})]\bullet[0.3\ 0.5\ 0.1\ 0.2.1\ 3],$$

where $\alpha_{pr}$ = the variance of elevation along the ray; $\beta_{pr}$ = the percent of non-road patches in ray; $\gamma_{pr}$ = distance from the ray's tip to dest; and $\delta_{pr}$ = distance of the ray's tip from the line between start and dest; $\varepsilon_{pr}$ and $\zeta_{pr}$ have the same definitions as they do in the urban mode.

A developer continues to build a primary road until it either connects with the existing primary road network, or it reaches the edge of the world. Before committing the road, the developer clips it if it extends too far over water, and runs a smoothing filter over it to remove jagged angles.

## 6 Results

Figure 1 shows a vectorized version of the output of one of our simulations in a world of roughly nine square miles. Here the artist has brought almost all commercial (red) development to the bay with a strong dose of honey. Note in particular the four differently gridded road layouts, surrounded by ungridded, organic layouts. Vectorization is a polygon- and spline-based 2D post-process we apply to the raster outlines of parcels and roads in our simulation output.

Figure 15 shows the developmental history of another city. Initial parameters including mountains, water, and road layouts were painted onto the map in less than 10 minutes. All development thereafter was completely automated. Note in particular the automatic growth of primary roads and the development focused along primary roads and shorelines.

Validating this input numerically is a challenge because we are not attempting to reproduce any existing city, but only the typical urban patterns that make a newly visited city feel "real". To address this problem, urban geographers have turned to measures of *composition* or proportions of land cover by use; and *spatial configuration*, which refers to the distribution and clustering of land use [Turner 1989; Torrens 2002, 2003]. As we have demonstrated, composition is easily controlled by the artist, and proportions similar to those mentioned by Eisner et al. [1993] are simple to produce. Spatial configuration can be measured with conditional probabilities. We call our measure *conditional proportion* ($C_p$), inspired by the urban geography literature:

$$C_p = \frac{\sum_{i=1}^{m}\sum_{j=1}^{n}\left(UseIs(p_{i,j},u) \times \sum_{k=i-r}^{i+r}\sum_{l=j-r}^{j+r} UseIs(p_{k,l},v)\right)}{\sum_{i=1}^{m}\sum_{j=1}^{n}\left(UseIs(p_{i,j},u) \times \sum_{k=i-r}^{i+r}\sum_{l=j-r}^{j+r} InMap(p_{k,l})\right)}.$$

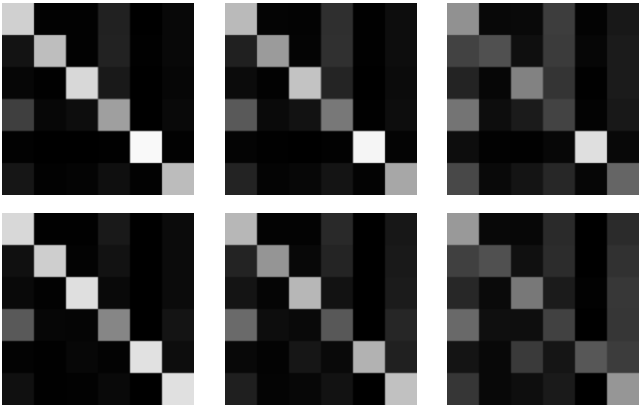

**Figure 16:** Strikingly similar spatial configurations of use between Houston, TX (bottom row) and our simulation (top row). Simulation results are the average configuration of 20 unsteered outputs. The left column of images depicts configuration in a 40 ft radius, the middle in a 160 ft radius, and the right in a 640 ft radius. Rows and columns within images index the residential, commercial, industrial, road, water and other uses (left to right, top to bottom). Each cell in an image indicates the likelihood that the column's use will be in the radius surrounding the row's use, with white indicating high likelihood. Note the strong diagonal indicating clustering by use, and the proximity of roads to property at the highest radius.

$C_p$ is the proportion of land within a radius $r$ surrounding use $u$ that is dedicated to use $v$, across the entire $m \times n$ map. Here $i, j, k$ and $l$ are map coordinates, $p_{i,j}$ is the patch at $i,j$, $UseIs(p_{i,j},u)$ is a function that returns 1 if $p_{i,j}$ has use $u$ and 0 otherwise, $InMap(p_{i,j})$ is a function that returns 1 if $p_{i,j}$ is in the map and 0 otherwise.

We use $C_p$ to compare our output to Houston, TX and show the results in Figure 16. Here we visualize conditional relationships between each type of use with grayscale matrices. In each matrix, the row indicates use $u$, the column use $v$, with the gray level in each cell indicating the corresponding value of $C_p$ (brighter values are higher). Note in particular the bright diagonals, which indicate self-clustering of various usage types. Two other strong trends are the proximity of roads to most types (indicated by the bright column, fourth from left in each matrix), and the fading of strong pattern with increasing scale.

The 2D maps produced by our system are designed to be populated with 3D building and structure models, providing urban content for digital productions. Figure 17 illustrates the potential of our maps for this application by visualizing one of them using Electronic Arts' SimCity 3000 [Electronic Arts] display engine.

Clearly our maps include enough information to enable meaningful choices among SimCity's library of buildings. Our maps add an interesting ungridded and higher level structure not normally present in SimCity's output. Note that because SimCity 3000 cannot not display primary roads, place any structures on inclines, nor easily fill non-rectangular parcels, we were forced to simplify our output considerably.

## 7 Limitations and future work

Although already quite useful, our simulation does have some limitations. Our largest nine square mile worlds take several hours to develop fully. These sorts of speeds are far less than interactive and must be improved before our tool can see practical use. However, to this point our focus has been on correctness and robustness, not speed, and optimizations should be simple to implement. For example, NetLogo itself is a highly transparent and interruptible programming language and environment designed for instruction, and implemented in Java. Implementing the simulation directly in a high-performance language such as C++ would be an obvious first step.

Of course, there is a great deal of urban development itself that our simulation, like any, does not capture. We are not completely satisfied with the shape of parcel boundaries even after vectorization – real cities often show much more regularity in their parceling. We might capture some of this regularity by performing development or at least parceling in larger units: block by block, or tract by tract, much like civil zoning authorities and modern developers. In addition, in the real world, uses are mixed even *within* parcels. Mechanisms for mixed-type developments would be a useful addition. With only two levels to our transport hierarchy, our cities often look a bit "flat", without the rich structure that a deep transport hierarchy brings. A new level such as freeways would be welcome.

In the longer run, we face important and exciting challenges. Although we already offer good artist control of the simulation, control at higher historic or cultural levels (e.g. "old town" or "Chinatown") would increase the value of our tool greatly. To achieve this goal, we will have to simulate changes in urban dynamics over time: for example, the changes in transport from horses, to trains, to cars.

## 8 Acknowledgements

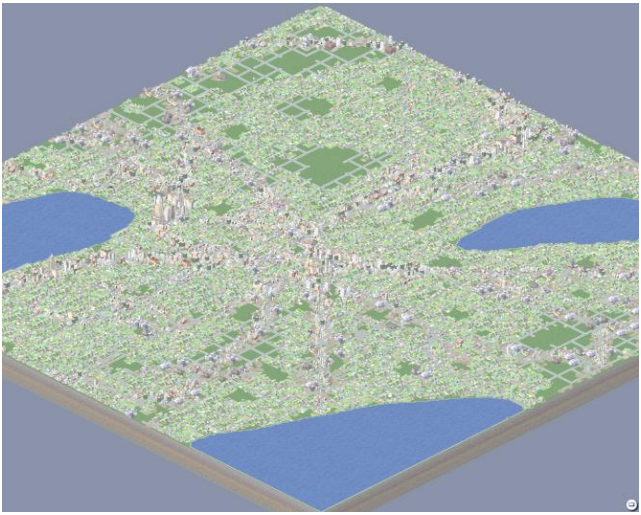 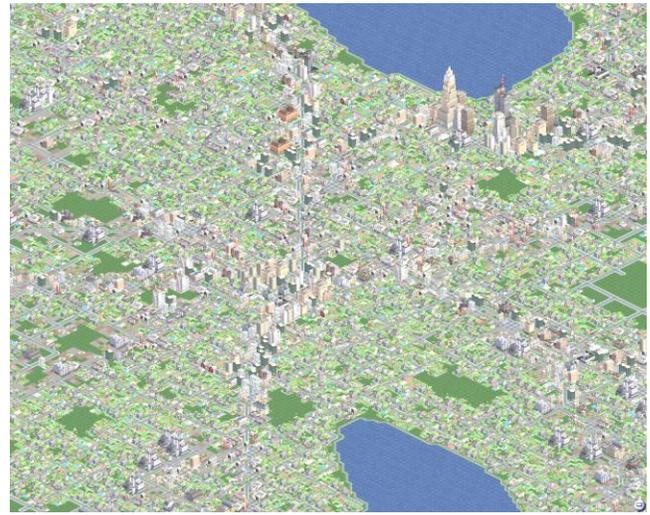

**Figure 17:** Visualization of simulation results using SimCity 3000. Left, an overview shows a curvilinear, hierarchical road network and two urban cores. Right, a close-up on the cores.